\documentclass[aps,twocolumn,prd,showpacs,showkeys,preprintnumbers,superscriptaddress,nofootinbib,floatfix,longbibliography]{revtex4-2}[prd]

\usepackage{orcidlink}
\usepackage{lipsum}
\usepackage{amsmath}
\usepackage{amsfonts}
\usepackage{amssymb}
\usepackage{mathrsfs}
\usepackage{graphicx}
\usepackage{color}
\usepackage[dvipsnames]{xcolor}
\usepackage{bm}
\usepackage{blindtext}
\usepackage{wasysym}
\usepackage{hyperref}
\hypersetup{colorlinks=true,allcolors=blue}

\usepackage[normalem]{ulem}
\usepackage{fontawesome} % for \faGithub or \faGithubSquare
\usepackage[all]{hypcap}
\usepackage{lineno}
\newcommand{\ie}{{\it i.e.}}

\newcommand{\bi}{\begin{itemize}}
	\newcommand{\ei}{\end{itemize}}

\begin{document}
	
	\title{Stringent constraints on nonstandard neutrino interactions using high-purity $\nu_{\mu}$ CC events in IceCube DeepCore~\href{https://github.com/JKrishnamoorthi/NSI-Limits-from-IceCube-DC-Public-Data}{\faGithubSquare}}
	
	\author{J Krishnamoorthi\,\orcidlink{0009-0006-1352-2248}}
	\affiliation{Institute of Physics, Sachivalaya Marg, Sainik School Post, Bhubaneswar 751005, India}
	\affiliation{Department of Physics, Aligarh Muslim University, Aligarh 202002, India}
	
	\author{Anil Kumar\,\orcidlink{0000-0002-8367-8401}}
	\affiliation{Institute of Physics, Sachivalaya Marg, Sainik School Post, Bhubaneswar 751005, India}
	
	\author{Sanjib Kumar Agarwalla\,\orcidlink{0000-0002-9714-8866}}
	\affiliation{Institute of Physics, Sachivalaya Marg, Sainik School Post, Bhubaneswar 751005, India}
	\affiliation{Homi Bhabha National Institute, Training School Complex, Anushakti Nagar, Mumbai 400094, India\\
	{\tt krishnamoorthi.j@iopb.res.in, anil.k@iopb.res.in, sanjib@iopb.res.in} \smallskip}

	\preprint{IOP/BBSR/2025-05}
	
	\date{\today}
	
	\begin{abstract}
		
		The neutral-current (NC) nonstandard interactions (NSIs) of neutrinos with fermions can modify the flavor oscillations of atmospheric neutrinos as they propagate through the Earth. We present constraints on the NC-NSI parameters $\varepsilon_{\mu\tau}$ and $\varepsilon_{\tau\tau}-\varepsilon_{\mu\mu}$ (one at a time) using a high-purity sample of $\nu_{\mu}$ charged-current atmospheric neutrino events collected by IceCube DeepCore over 7.5 years of livetime. These two parameters significantly affect the $\nu_\mu$ disappearance channel for which this golden event sample is optimized by IceCube Collaboration. The best fit to this dataset is consistent with no NSI hypothesis, and we place the stringent constraints $-\,0.0094 < \varepsilon_{\mu\tau} < 0.0079$ and $-\,0.031 < \varepsilon_{\tau\tau}-\varepsilon_{\mu\mu} <  0.029$ at the 90\% confidence level, assuming normal mass ordering.
		
	\end{abstract}
	
	% \pacs{XXXX}
	% \keywords{XXXX}
	
	\maketitle
	
	%%%%%%%%%%%%%%%%%%%%%%%%%%%%%%%%%%%%%%%%%%%%%%
	\section{Introduction}
	The Nobel Prize in Physics (2015) was awarded for the discovery of neutrino oscillations~\cite{Super-Kamiokande:1998kpq, SNO:2001kpb, SNO:2002tuh}, which provided direct evidence of non-zero neutrino mass, thereby requiring theories beyond the Standard Model of particle physics. As we celebrate the tenth anniversary of this Nobel Prize, measurements of neutrino oscillation parameters have now entered into a precision era~\cite{ParticleDataGroup:2024cfk,Esteban:2024eli,Capozzi:2025ovi}. Over this decade, IceCube DeepCore~\cite{IceCube:2011ucd} has emerged as a prominent atmospheric neutrino experiment, complementing the achievements of Super-Kamiokande~\cite{Super-Kamiokande:2002weg}. With improved precision measurements of oscillation parameters, neutrinos provide a unique platform to search for new physics scenarios.
	
	Considering the Standard Model (SM) of particle physics as an effective field theory~\cite{Weinberg:1967tq} allows us to introduce the dimension-6 operator~\cite{Gavela:2008ra, Bischer:2019ttk} corresponding to new interactions of neutrinos beyond the SM. These nonstandard interactions (NSIs) of neutrinos were originally proposed by Wolfenstein~\cite{Wolfenstein:1977ue} as a possible underlying mechanism behind neutrino flavor transitions. NSIs could happen via charged current (CC) or neutral current (NC), which have been studied extensively in the literature~\cite{Valle:1987gv,Roulet:1991sm, Guzzo:1991hi,Guzzo:2000kx,Huber:2001zw,Gonzalez-Garcia:2004pka,Kopp:2008,Biggio:2009nt,Escrihuela:2011cf,Gonzalez-Garcia:2011vlg,Agarwalla:2012wf,Ohlsson:2012kf,Esmaili:2013fva,Gonzalez-Garcia:2013usa,MINOS:2013hmj,Chatterjee:2014gxa,Agarwalla:2014bsa,Mocioiu:2014gua,Miranda:2015dra,Agarwalla:2015cta,Choubey:2015xha,Agarwalla:2016fkh,Salvado:2016uqu,Farzan:2017xzy,Coloma:2017egw,IceCube:2017zcu,Esteban:2018ppq,Borexino:2019mhy,Bhupal:2019qno,Khatun:2020,Kumar:2021lrn,Agarwalla:2021zfr,IceCubeCollaboration:2021euf,IceCube:2022pbe,Krishnamoorthi:2025,Coloma:2023ixt,NOvA:2024lti,IceCubeCollaboration:2021euf,KM3NeT:2024pte,ANTARES:2021crm, IceCube:2022ubv,Super-Kamiokande:2011dam,Jana:2024lfm}. The CC-NSIs affect the neutrino production and the detection, whereas the NC-NSIs affect the neutrino propagation through matter.  In this work, we utilize oscillations of atmospheric neutrinos passing through Earth to constrain the NC-NSIs. 
	
	The effects of the NC-NSIs can be incorporated in the neutrino oscillations by modifying the matter potential with the effective Hamiltonian in flavor basis, given as
	\begin{align}
		H_{\rm eff} &= \frac{1}{2E} \, U \left(\begin{array}{ccc}
		0 & 0 & 0 \\
		0 & \Delta m^2_{21} & 0 \\
		0 & 0 & \Delta m^2_{31}\end{array}
		\right) U^\dag \nonumber \\
		& + 
		V_{\rm CC} \left(\begin{array}{ccc}
			1 +~\varepsilon_{ee} - \varepsilon_{\mu\mu} & \varepsilon_{e\mu} & \varepsilon_{e\tau} \\ \varepsilon_{e\mu}^* & 0 & \varepsilon_{\mu\tau}\\ 
			\varepsilon_{e\tau}^* & \varepsilon_{\mu\tau}^* & \varepsilon_{\tau\tau} - \varepsilon_{\mu\mu}
		\end{array}
		\right),
		\label{eq:modified_H}
	\end{align}	
	where $U$ is the Pontecorvo-Maki-Nakagawa-Sakata matrix~\cite{Maki:1962mu}, $\Delta m^2_{ij} \equiv m_i^2 - m_j^2$ are the mass-squared differences, and $V_{\rm CC} \equiv \sqrt{2}G_F N_e$~\cite{Opher:1974drq, Langacker:1982ih} represents the standard interaction (SI) potential due to the charged-current coherent forward scattering of neutrinos with the ambient electrons with number density $N_e$. The parameters $\varepsilon_{\alpha\beta}$ are the NSI coupling parameters defined as $\varepsilon_{\alpha\beta}=\varepsilon_{\alpha\beta}^e+3\varepsilon_{\alpha\beta}^u+3\varepsilon_{\alpha\beta}^d$, representing contributions from interactions with electrons, up quarks, and down quarks, respectively. The diagonal parameters are real valued, whereas the off-diagonal parameters can be complex \ie, $\varepsilon_{\alpha\beta} \equiv |\varepsilon_{\alpha\beta}| e^{i\phi_{\alpha\beta}}$, with $\varepsilon_{\alpha\beta}^*$ denoting their complex conjugates. For antineutrinos,  $U \rightarrow U^*$, $V_{\rm CC} \rightarrow -V_{\rm CC}$, and $\varepsilon_{\alpha\beta} \to \varepsilon_{\alpha\beta}^*$.

	\begin{figure}[htp!]
		\includegraphics[width=\linewidth]{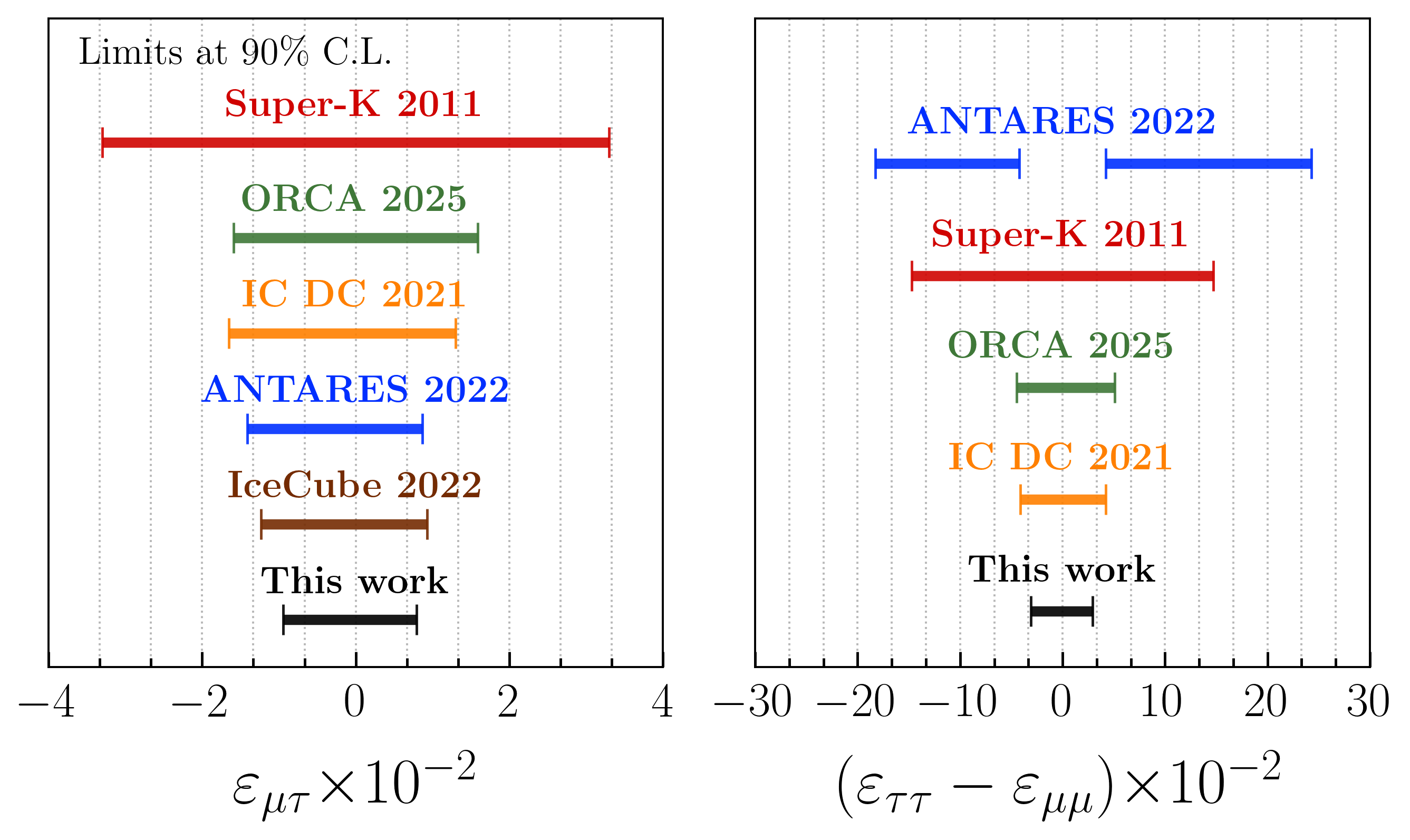}
		\caption{Constraints on the NSI parameters $\varepsilon_{\mu\tau}$ and $\varepsilon_{\tau\tau}-\varepsilon_{\mu\mu}$ at the 90\% confidence level (CL) from this analysis using the eight-year golden event sample of IceCube DeepCore. Comparison with limits from other experiments, such as IceCube DeepCore (2021)~\cite{IceCubeCollaboration:2021euf}, KM3NeT/ORCA (2025)~\cite{KM3NeT:2024pte}, ANTARES (2022)~\cite{ANTARES:2021crm}, IceCube (2022)~\cite{IceCube:2022ubv}, and Super-Kamiokande (2011)~\cite{Super-Kamiokande:2011dam}, is shown. The limits for $\varepsilon_{\mu\tau}$ are obtained by fixing the complex phase to $0$ (\ie, positive $\varepsilon_{\mu\tau}$) and $\pi$ (\ie, negative $\varepsilon_{\mu\tau}$). For consistency, bounds from Refs.~\cite{KM3NeT:2024pte,ANTARES:2021crm, IceCube:2022ubv,Super-Kamiokande:2011dam} have been rescaled to the NSI convention used in our work.}
		\label{fig:nsi_comparison}
	\end{figure}
	
	In this work, we constrain the NC-NSI parameters $\varepsilon_{\mu\tau}$ and $\varepsilon_{\tau\tau}-\varepsilon_{\mu\mu}$ considering one at a time, which significantly affect the $\nu_\mu$ survival probability, $P(\nu_\mu \rightarrow \nu_\mu)$. This $\nu_\mu$ disappearance channel is mainly governed by the atmospheric neutrino oscillation parameters $\Delta m_{31}^2$ and $\theta_{23}$, where $\Delta m_{31}^2$ affects the location of the oscillation valley, and $\theta_{23}$ affects the depth of the oscillation valley. The NSI parameters $\varepsilon_{\mu\tau}$ and $\varepsilon_{\tau\tau}-\varepsilon_{\mu\mu}$ also affect the location and depth of the oscillation valley, which indicates their correlations with $\Delta m_{31}^2$ and $\theta_{23}$. This makes the $\nu_\mu \rightarrow \nu_\mu$ channel more suitable for probing these two parameters. To probe these two NSI parameters, we use a publicly available atmospheric neutrino dataset~\cite{DVN_B4RITM_2025} from IceCube DeepCore with 7.5 years of livetime that is specifically optimized for $\nu_\mu$ CC events. This high-purity $\nu_\mu$ CC sample has an excellent sensitivity to the $\nu_\mu$ disappearance channel. Therefore, this sample is expected to probe the NSI parameters  $\varepsilon_{\mu\tau}$ and $\varepsilon_{\tau\tau}-\varepsilon_{\mu\mu}$ efficiently.

	The remaining NSI parameters $\varepsilon_{e\mu}$, $\varepsilon_{e\tau}$, and $\varepsilon_{ee}-\varepsilon_{\mu\mu}$ have been studied in our other work~\cite{Krishnamoorthi:2026zve} using the same data sample that we use in the present work. These NSI parameters are not expected to significantly modify the location and depth of the oscillation valley in the $\nu_\mu$ disappearance channel, rather, they mainly alter the $\theta_{13}$-driven Earth's matter effects. The sensitivity to $\varepsilon_{e\mu}$ and $\varepsilon_{e\tau}$ parameters is primarily driven by the appearance channels, which are subject to degeneracies involving the Dirac-{$CP$} phase $\delta_{CP}$. Since the data sample is dominated by $\nu_\mu\rightarrow\nu_\mu$ disappearance events, for which $\delta_{CP}$  dependence is strongly suppressed by the factor $(\Delta m_{21}^2/\Delta m_{31}^2)\times\sin\theta_{13}\sim 0.005$, the constraints obtained on $\varepsilon_{e\mu}$ and $\varepsilon_{e\tau}$ parameters are free of $\delta_{CP}$ degeneracy. These $\delta_{CP}$-free constraints are comparable and complementary to the existing limits from long-baseline experiments.
	
	In the present work, while probing $\varepsilon_{\mu\tau}$ and $\varepsilon_{\tau\tau}-\varepsilon_{\mu\mu}$, we assume normal mass ordering (NO). An analysis under the inverted mass ordering (IO) hypothesis is not possible because the detector-related systematic uncertainty treatment is available only for the NO configuration in the public IceCube DeepCore data release. However, the impact of mass ordering uncertainty on the NSI sensitivities has been quantified by excluding detector-related systematic uncertainties and the effect is found to be small. The details of this study are discussed in Appendix~\ref{app:io_sensitivity}. Figure~\ref{fig:nsi_comparison} presents updated constraints on the NSI parameters $\varepsilon_{\mu\tau}$ and $\varepsilon_{\tau\tau}-\varepsilon_{\mu\mu}$ obtained using this eight-year golden event sample of IceCube DeepCore, and a comparison is done with the limits from the other neutrino oscillation experiments.	Here, we treat $\varepsilon_{\mu\tau}$ as a real parameter while deriving the corresponding limits. The constraints derived in this work represent the stringent bounds assuming NO. For comparison, we have included the results from several neutrino oscillation experiments like IceCube DeepCore~\cite{IceCubeCollaboration:2021euf}, ORCA~\cite{KM3NeT:2024pte}, ANTARES~\cite{ANTARES:2021crm}, IceCube~\cite{IceCube:2022ubv}, and Super-Kamiokande~\cite{Super-Kamiokande:2011dam}.

	The previous IceCube DeepCore analysis~\cite{IceCubeCollaboration:2021euf} placed all-flavor constraints on NSI parameters using a three-year atmospheric neutrino dataset in the GeV-energy range, which was not specifically optimized for $\nu_\mu$ CC tracklike events, and had a total event sample more than twice that of the present analysis. In contrast, the present study employs a high-purity $\nu_\mu$ CC sample optimized for the $\nu_\mu$ disappearance channel enabling us to set the stringent constraints on $\varepsilon_{\mu\tau}$ and $\varepsilon_{\tau\tau}-\varepsilon_{\mu\mu}$.
	 
	At high energies above a few hundred GeV, standard neutrino oscillations are negligible and, therefore, the impact of $\varepsilon_{\mu\tau}$ on the $\nu_\mu$ disappearance channel appears as a clean signal. By focusing on this channel at TeV energies, IceCube Collaboration performed an analysis to set tight limits on $\varepsilon_{\mu\tau}$~\cite{IceCube:2022ubv}. On the other hand, the present study probes NSIs using low-energy DeepCore data in the oscillation-dominated GeV energy regime, where the NSI effects are subleading. Despite this more challenging scenario, the present analysis with an optimized data sample having high-quality $\nu_\mu$ CC events provides the strongest constraints on $\varepsilon_{\mu\tau}$ and $\varepsilon_{\tau\tau}-\varepsilon_{\mu\mu}$ irrespective of the current uncertainties on various oscillation and systematic parameters.
	
	In other words, our study shows that a well-optimized data sample tailored for a specific physics question can outperform analyses using larger but less targeted datasets. Our analysis procedure to obtain the above-mentioned constraints is described in detail in the later part of the article. Let us first understand how the presence of NSI affects the neutrino oscillation patterns. \\
	
	%%%%%%%%%%%%%%%%%%%%%%%%%%%%%%%%%%%%%%%%%%%%%%%

	%%%%%%%%%%%%%%%%%%%%%%%%%%%%%%%%%%%%%%%%%%%%%%%%
	\section{Effects of NSI on neutrino oscillations}
	The presence of NSIs modifies the effective matter potential experienced by neutrinos as they propagate through the Earth. This can lead to significant alterations in the oscillation probabilities, depending on the magnitude and sign of the NSI parameters. The following Eq.~(\ref{eqn:theta23_evol}) taken from the Ref.~\cite{Agarwalla:2021zfr} shows the modified mixing angle $\theta_{23}$ in matter due to the presence of NSI parameters $\varepsilon_{\mu\tau}$ and $\varepsilon_{\tau\tau}-\varepsilon_{\mu\mu}$:
    \begin{equation}
    \tan 2\theta_{23}^m \simeq 
    \frac{(c_{13}^2 - \alpha c_{12}^2)\sin 2\theta_{23} + 2\varepsilon_{\mu\tau}\hat{A}}
    {(c_{13}^2 - \alpha c_{12}^2)\cos 2\theta_{23} + (\varepsilon_{\tau\tau} - \varepsilon_{\mu\mu})\hat{A}}\,,
    \label{eqn:theta23_evol}
    \end{equation}
    where $\alpha=\Delta m_{21}^2/\Delta m_{31}^2$ and $\hat{A}=V_{CC}/\Delta m_{31}^2$. Considering one NSI parameter at a time, both $\varepsilon_{\mu\tau}$ and $\varepsilon_{\tau\tau}-\varepsilon_{\mu\mu}$ will be having an impact on the mixing angle $\theta_{23}$, which could affect the probability amplitude. In addition, the NSI parameter $\varepsilon_{\mu\tau}$ also modifies the parameter $\Delta m_{31}^2$~\cite{Agarwalla:2021zfr}, which results in a significant change in the location of the oscillation valley.

    \begin{figure}[htp!]
        \includegraphics[width=\linewidth]{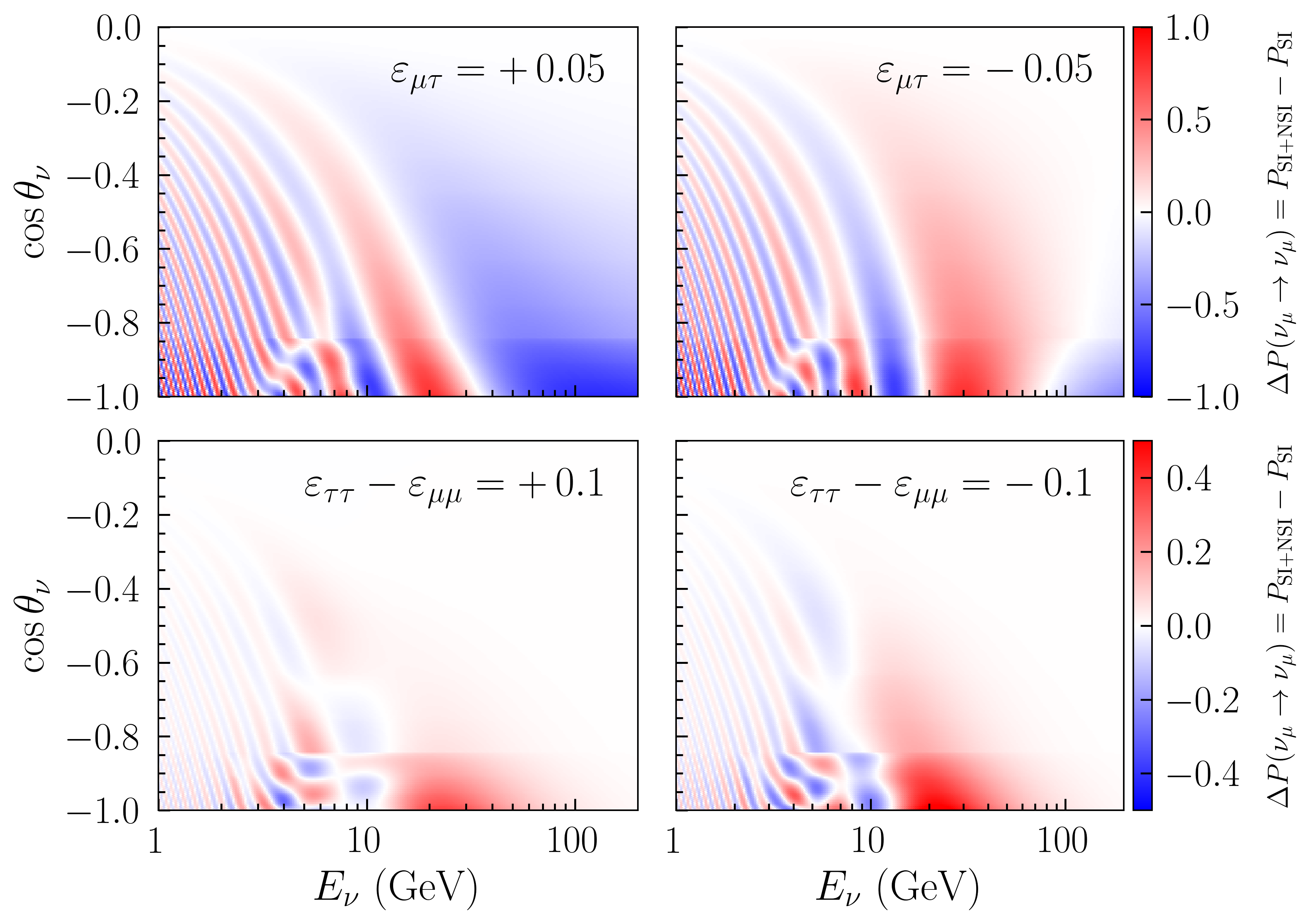}
        \caption{Difference in the three-flavor neutrino oscillation probabilities $P(\nu_\mu \rightarrow \nu_\mu)$ between SI + NSI and SI scenarios in the plane of neutrino energy and cosine of zenith angle. The top and bottom panels correspond to the NSI scenarios with $\varepsilon_{\mu\tau} = \pm\,0.05$ and $\varepsilon_{\tau\tau} - \varepsilon_{\mu\mu} = \pm\,0.1$, respectively, considering one NSI parameter at a time. Here, we take $\theta_{23} = 45.57^\circ$ and $\Delta m^2_{31} = 2.48 \times 10^{-3}~{\rm eV}^2$.}
        \label{fig:fv_nsi_prob}
    \end{figure}

    Figure~\ref{fig:fv_nsi_prob} shows the effects of the NSI parameters $\varepsilon_{\mu\tau}$ and $\varepsilon_{\tau\tau}-\varepsilon_{\mu\mu}$ one at a time in the plane of neutrino energy and the cosine of zenith angle. The difference of probabilities $P(\nu_{\mu}\rightarrow\nu_{\mu})$ between SI + NSI and SI scenarios shows that the presence of $\varepsilon_{\mu\tau} = \pm \, 0.05$ shifts the location of the oscillation valley, which results in the prominent alternative patches of blue and red. Additionally, the direction of shift in the oscillation valley depends on the sign of $\varepsilon_{\mu\tau}$. On the other hand, the presence of $\varepsilon_{\tau\tau}-\varepsilon_{\mu\mu} = \pm\,0.1$ mainly affects the mixing angle $\theta_{23}$ and results in a modification in the depth of the oscillation valley. Note that the direction of this modification is independent of the sign of $\varepsilon_{\tau\tau}-\varepsilon_{\mu\mu}$. Comparing the effects of $\varepsilon_{\mu\tau}$ and $\varepsilon_{\tau\tau}-\varepsilon_{\mu\mu}$ on the oscillograms, the parameter $\varepsilon_{\mu\tau}$ has an impact up to high-energy regions, meanwhile the parameter $\varepsilon_{\tau\tau}-\varepsilon_{\mu\mu}$ shows an effect only below 100 GeV. To probe these two NSI parameters, we utilize a publicly available  atmospheric neutrino data sample~\cite{DVN_B4RITM_2025} collected during 2011--2019 by IceCube DeepCore.\\
	
	%%%%%%%%%%%%%%%%%%%%%%%%%%%%%%%%%%%%%%%%%%%%%%%%%
	\section{Neutrino events at IceCube DeepCore}
	IceCube~\cite{IceCube:2016zyt} is a neutrino observatory located deep inside the Antarctic ice at the South Pole. The detector consists of 5,160 digital optical modules (DOMs) deployed along 86 vertical strings below about 1.5 km of ice. The DOMs detect the Cherenkov photons emitted by the secondary charged particles that are produced during the interactions of neutrinos with ice. The $\nu_{\mu}$ CC interactions produce muons which deposit energies in the form of long tracks, whereas $\nu_{e}$ CC, $\nu_{\tau}$ CC, and the NC interactions of neutrinos of all flavors deposit energies in the form of cascades. Some $\nu_{\tau}$ CC events may also give rise to muons from the tau decay, resulting in tracks. Since the neutrino and antineutrino events have similar topologies, we perform our analysis without distinguishing them.
	
	DeepCore is a densely instrumented central region of IceCube having DOMs with higher quantum efficiency. The reduced spacing between the neighboring DOMs in DeepCore, along with the enhanced DOM efficiency, helps DeepCore in detecting lower-energy neutrinos in the multi-GeV range. In the present analysis, we use the eight-year data sample of atmospheric neutrinos of high-purity $\nu_{\mu}$ CC events observed at DeepCore in the reconstructed energy range of 6.3 GeV to 158.5 GeV~\cite{DVN_B4RITM_2025}. This dataset, consisting of 21,914 events, corresponds to a ``golden event sample,'' where only those events are reconstructed that consist of direct photon hits without significant scattering in ice. Also, this sample is provided with the Monte Carlo (MC) events simulated using the full IceCube simulation chain~\cite{IceCube:2019dqi}.  Each event can be weighted to take into account the atmospheric neutrino flux~\cite{Honda:2015fha}, neutrino interaction cross section in the detector, three-flavor neutrino oscillations, and detector response such that the simulated sample accurately matches the experimentally observed data.
	
	This golden event sample has been updated with the improved simulation, calibration, selection techniques, and reconstruction algorithms as described in detail in Ref.~\cite{IceCubeCollaboration:2023wtb}. The selection criteria, filters, and reconstruction algorithms are developed using the simulated events, and the same are applied to the observed data to perform the analysis. Various kinds of filters have been applied to suppress the backgrounds like cosmic muons and detector noise, below 1\% to obtain the neutrino-dominated sample.
	
	Reconstruction algorithms based on the maximum likelihood technique~\cite{IceCube:2014flw,Garza2014Measurement,AndriiThesis,IceCube:2022kff} have been employed to reconstruct observables like energy ($E_{\rm reco}$) and direction ($\cos\theta_{\rm reco}$) of a neutrino event. A particle identification (PID) score for each neutrino event is estimated using a boosted decision tree (BDT)~\cite{Friedman:2001wbq}. The PID score represents the probability of an event having the tracklike topology. In this sample, the data are binned such that $E_{\rm reco}$ has 10 logarithmic bins in the range $6.3$ GeV to $158.5$ GeV and  $\cos\theta_{\rm reco}$ has 10 linear bins in the range $-1$ to $0.1$. To retain the sufficient statistics, the last energy bin is defined with twice the usual bin width. According to the event topologies, the events are further categorized into two PID bins: mixed events ($0.55 - 0.75$) and tracklike events ($0.75 - 1.0$). The cascadelike events with ${\rm PID} < 0.55$ have been removed by the Collaboration to obtain the high-purity $\nu_\mu$ CC sample, hence, those events are not present in the sample.
	
	%%%%%%%%%%%%%%%%%%%%%%%%%%%%%%%%%%%%%%%%%%%%%%%%%
	\begin{figure}[htp!]
		\includegraphics[width=\linewidth]{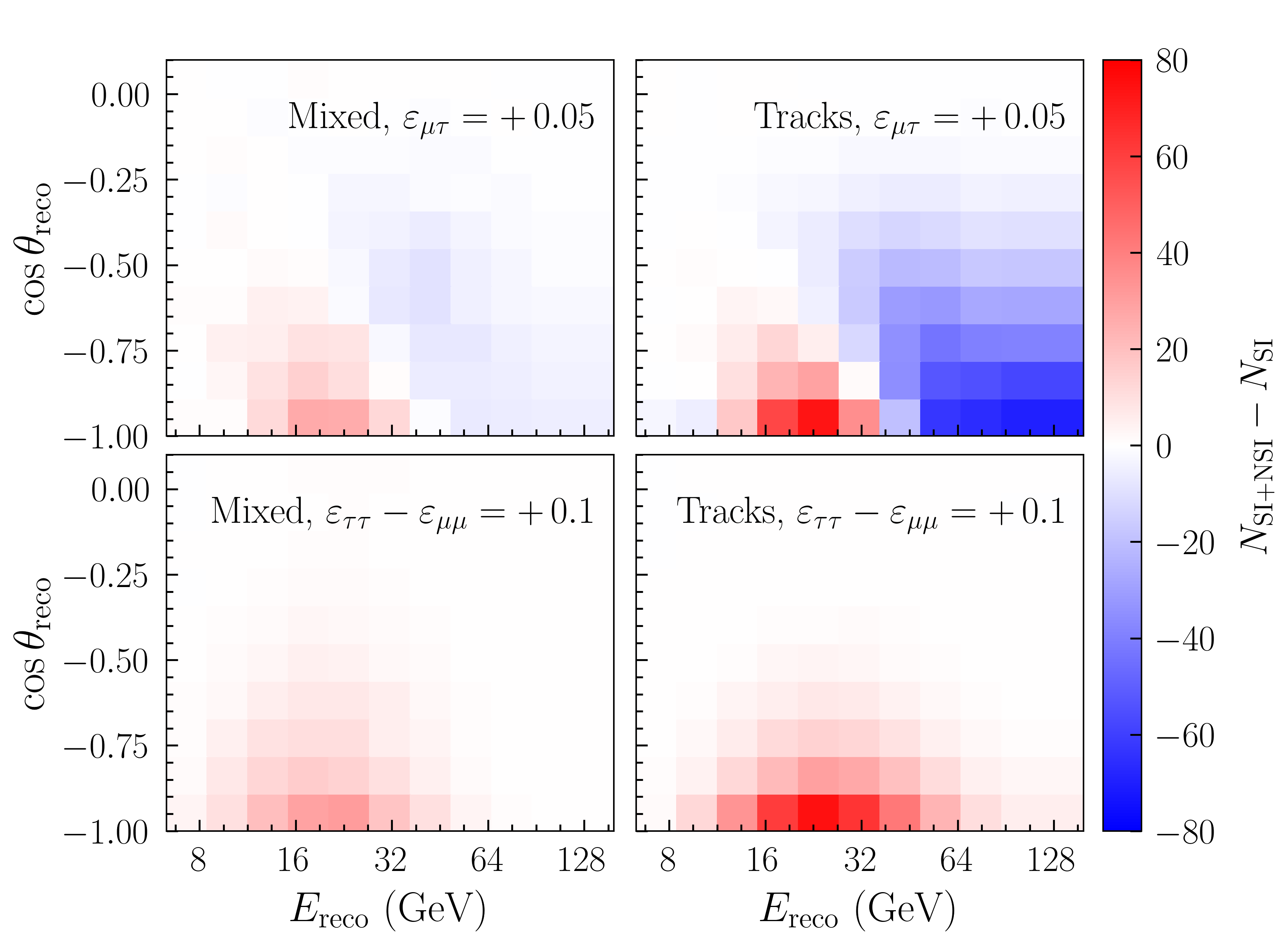}
		\caption{Difference of the expected event distributions between SI + NSI and SI scenarios (\ie,  $N_{\text{SI+NSI}} - N_{\text{SI}}$) in the plane of the reconstructed energy $E_{\text{reco}}$ and the reconstructed cosine of zenith angle $\cos\theta_{\text{reco}}$, for two NSI parameter choices taken one at a time. The top and bottom panels correspond to $\varepsilon_{\mu\tau} = +\,0.05$ and $\varepsilon_{\tau\tau} - \varepsilon_{\mu\mu} = +\,0.1$, respectively. The left and right panels show distributions for mixed and tracklike samples, respectively. Here, we use the nominal values of the nuisance parameters as given in Table~\ref{tab:systematic_params}.
		}
		\label{fig:event_dif_pos}
	\end{figure}
	
	To study the effects of NSIs on expected events at DeepCore, we show the distributions of event difference between SI + NSI and SI scenarios in Fig.~\ref{fig:event_dif_pos}. These distributions indicate the regions in reconstructed energy and cosine zenith that contribute most to the sensitivity. In the top panels, we take the NSI with a benchmark value of $\varepsilon_{\mu\tau}=+\,0.05$, considering other NSI parameters to be zero, where the event differences occur with alternating signs around the oscillation valley up to the high-energy region. In the bottom panels, we take NSI with $\varepsilon_{\tau\tau}-\varepsilon_{\mu\mu}=+\,0.1$, considering other NSI parameters to be zero, where the event differences occur only in the positive direction. These features are consistent with those observed in the probability oscillograms on the left panels of Fig.~\ref{fig:fv_nsi_prob}, which demonstrate the fidelity of the detector in retaining the effects of NSIs on the oscillation patterns even after considering all neutrino flavors, interaction kinematics, and the detector response. The distributions of event difference for the negative values of these NSI parameters are discussed in Appendix~\ref{app:event_diff}.\\

	%%%%%%%%%%%%%%%%%%%%%%%%%%%%%%%%%%%%%%%%%%%%%%%%%%%% 
	
	%%%%%%%%%%%%%%%%%%%%%%%%%%%%%%%%%%%%%%%%%%%%%%%%%%%%%%%%%%%%%%%%%%%%%%
	\section{Numerical analysis}
	This analysis constrains the NSI parameters $\varepsilon_{\mu\tau}$ and $\varepsilon_{\tau\tau} - \varepsilon_{\mu\mu}$ using a binned $\chi^2$ approach, where each NSI parameter is constrained independently one at a time while keeping the other NSI parameters fixed to zero. The statistical analysis based on the Frequentist method is performed by calculating the binned $\chi^2$ between the observed and expected event counts in bins of reconstructed energy, cosine zenith, and PID. For this analysis, the test statistic is adapted from Ref.~\cite{IceCube:2017lak}
	\begin{equation} 
		\chi^2_{\rm mod} = \sum_{i \in {\rm bins}} \frac{(N_i^{\rm exp} - N_i^{\rm obs})^2}{N_i^{\rm exp} + (\sigma_i^{\rm sim})^2} + \sum_{j \in {\rm syst}} \frac{(s_j - \hat{s}_j)^2}{\sigma^2_{s_j}}\, , 
		\label{eqn:mod_chi2}
	\end{equation}
	where $N_i^{\rm exp}$ is the expected event count in the $i$th bin, which is the sum of the MC event weights. $N_i^{\rm obs}$ is the observed event counts in the $i$th bin. The term $\sigma_i^{\rm sim}$ accounts for the statistical uncertainty in the MC simulations that is used to calculate $N_i^{\rm exp}$. The second term is the penalty term to account for the prior information about the systematic uncertainties, where $s_j$ is the value of the $j$th systematic parameter, $\hat{s}_j$ is its nominal value, and $\sigma_{s_j}$ is the associated uncertainty. 
	
	The systematic uncertainties are incorporated in the analysis in the form of nuisance parameters by reweighting the expected MC events. The reweighting and the binned $\chi^2_{\rm mod}$ calculations are performed using the open-source analysis framework \texttt{PISA}~\cite{IceCube:2018ikn} provided by IceCube Collaboration. This analysis includes a total of 20 nuisance parameters, which are allowed to vary freely during the minimization of the $\chi^2_{\rm mod}$ function given in Eq.~(\ref{eqn:mod_chi2}). These nuisance parameters account for uncertainties arising from detector response, atmospheric neutrino flux, neutrino interaction cross section, atmospheric muon background, and the oscillation parameters $\theta_{23}$ and $\Delta m^2_{31}$ following the systematic treatment described in detail in Ref.~\cite{IceCubeCollaboration:2023wtb}. The remaining oscillation parameters are fixed to their best-fit values obtained from NuFit v5.2~\cite{Esteban:2020cvm}, except for $\delta_{CP}$, which is fixed to zero since its impact is negligible. The details of the systematic uncertainty parameters and their best-fit values obtained from the present analysis are provided in Appendix~\ref{app:systematic} (see Table~\ref{tab:systematic_params}).
	
	%%%%%%%%%%%%%%%%%%%%%%%%%%%%%%%%

	%%%%%%%%%%%%%%%%%%%%%%%%%%%%%%%
	\begin{figure}[htp!]
		\centering
		\includegraphics[width=\linewidth]{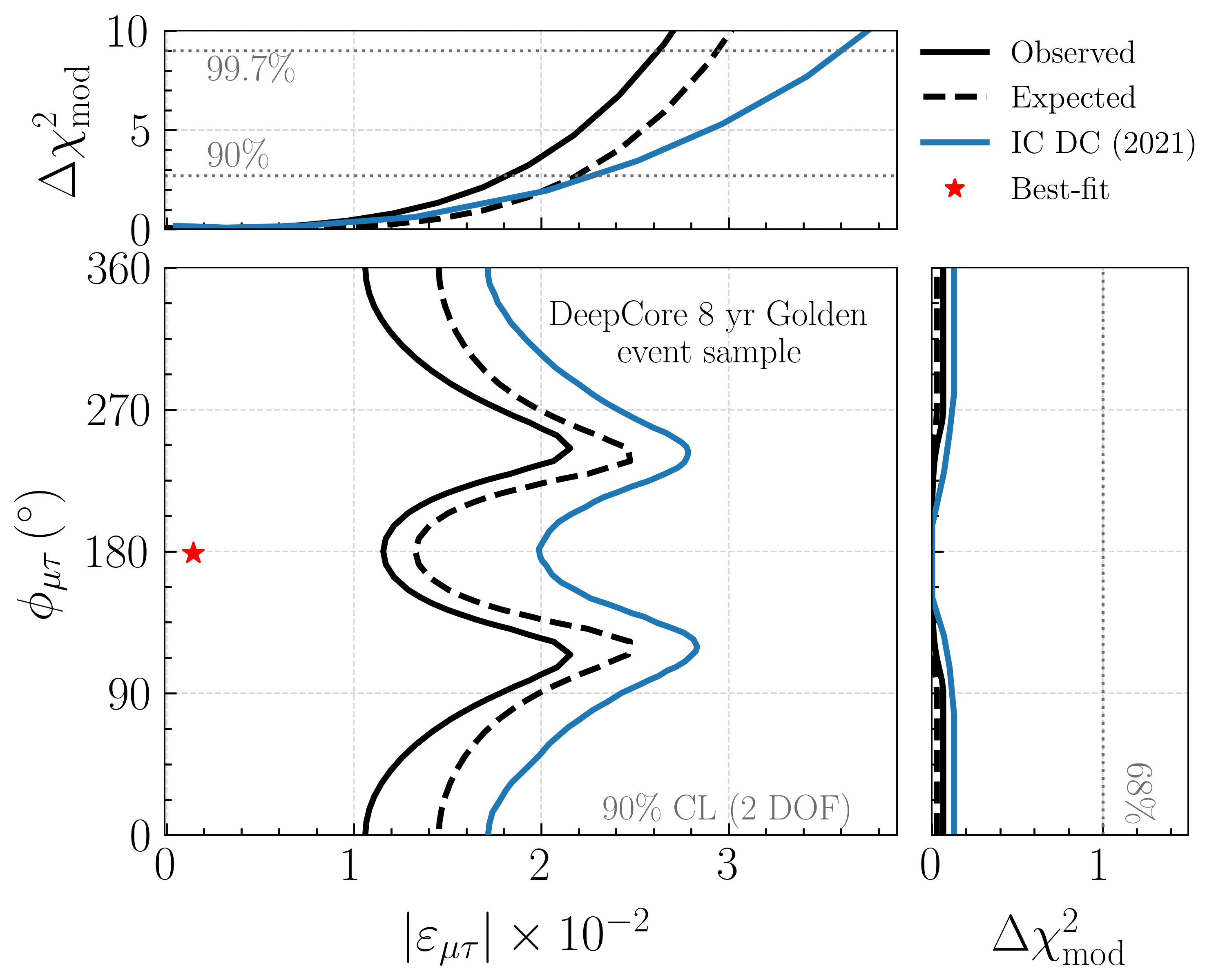}
		\caption{The observed (black solid curve) and the expected (black dashed curve) constraints on the magnitude $|\varepsilon_{\mu\tau}|$ and complex phase $\phi_{\mu\tau}$ at the 90\% CL (2 DOF) using the eight-year golden event sample of DeepCore. The red star marker represents the best-fit values for magnitude and phase. The top and the side subpanels represent the 1D projections of the 2D contours for  $|\varepsilon_{\mu\tau}|$ and $\phi_{\mu\tau}$, respectively, while the $\chi^2$ is minimized over the other parameter. The gray dotted lines in the subpanels represent the 68.3\%, 90\%, and 99.7\% confidence levels for 1 DOF. The solid blue curves represent the previous IceCube DeepCore result~\cite{IceCubeCollaboration:2021euf}.}
		\label{fig:off_diag_scan}
	\end{figure}
	%%%%%%%%%%%%%%%%%%%%%%%%%%%%%%%
	
	\section{Results}
	Figure~\ref{fig:off_diag_scan} shows the 90\% CL allowed regions considering 2 degrees of freedom (DOF) for the magnitude ($|\varepsilon_{\mu\tau}|$) and phase ($\phi_{\mu\tau}$) of the NSI parameter $\varepsilon_{\mu\tau}$ using the eight-year golden event sample. The best-fit values of the magnitude and phase after fitting the data are $0.0014$ and $178.8^\circ$, respectively, as shown by the red star. The $\Delta\chi^2_{\rm mod}$ for the best fit relative to the standard interaction is 0.06, indicating the data to be consistent with the standard interaction hypothesis with no NSI. The $p$-value of the NSI fit is 0.25, indicating good consistency of MC with the data (see Fig.~\ref{fig:data-mc} in Appendix~\ref{app:data_mc}). The black solid contour denotes the observed constraint at 90\% CL, while the black dashed contour represents the expected sensitivity by taking the best-fit values as the truth. For comparison, the contour from the previous IceCube DeepCore (2021) result~\cite{IceCubeCollaboration:2021euf} is also included as the blue curve. The observed and expected 90\% CL contours are in good agreement and are compatible with the previous IceCube DeepCore bounds~\cite{IceCubeCollaboration:2021euf}. The weak exclusion regions near $90^\circ$ and $270^\circ$ arise from the dependence of probability $P(\nu_\mu \rightarrow \nu_\mu)$ on the term $|\varepsilon_{\mu\tau}|\cos{\phi_{\mu\tau}}$~\cite{Kopp:2008}, which suppresses sensitivity in those regions. The top (side) subpanel displays the one-dimensional projection of the magnitude (phase), which is obtained by minimizing $\Delta\chi^2_{\rm mod}$ over phase (magnitude). Using the 1D projection, we obtain an upper bound of $|\varepsilon_{\mu\tau}| \leq 0.018$ at 90\% CL, which is the  stringent constraint on this parameter assuming NO. No constraint could be placed on the phase because the fitted magnitude goes to zero, resulting in no sensitivity to the phase.
	
	%%%%%%%%%%%%%%%%%%%%%%%%%%%%%%%%z
	\begin{figure}[htp!]
		\centering
		\includegraphics[width=\linewidth]{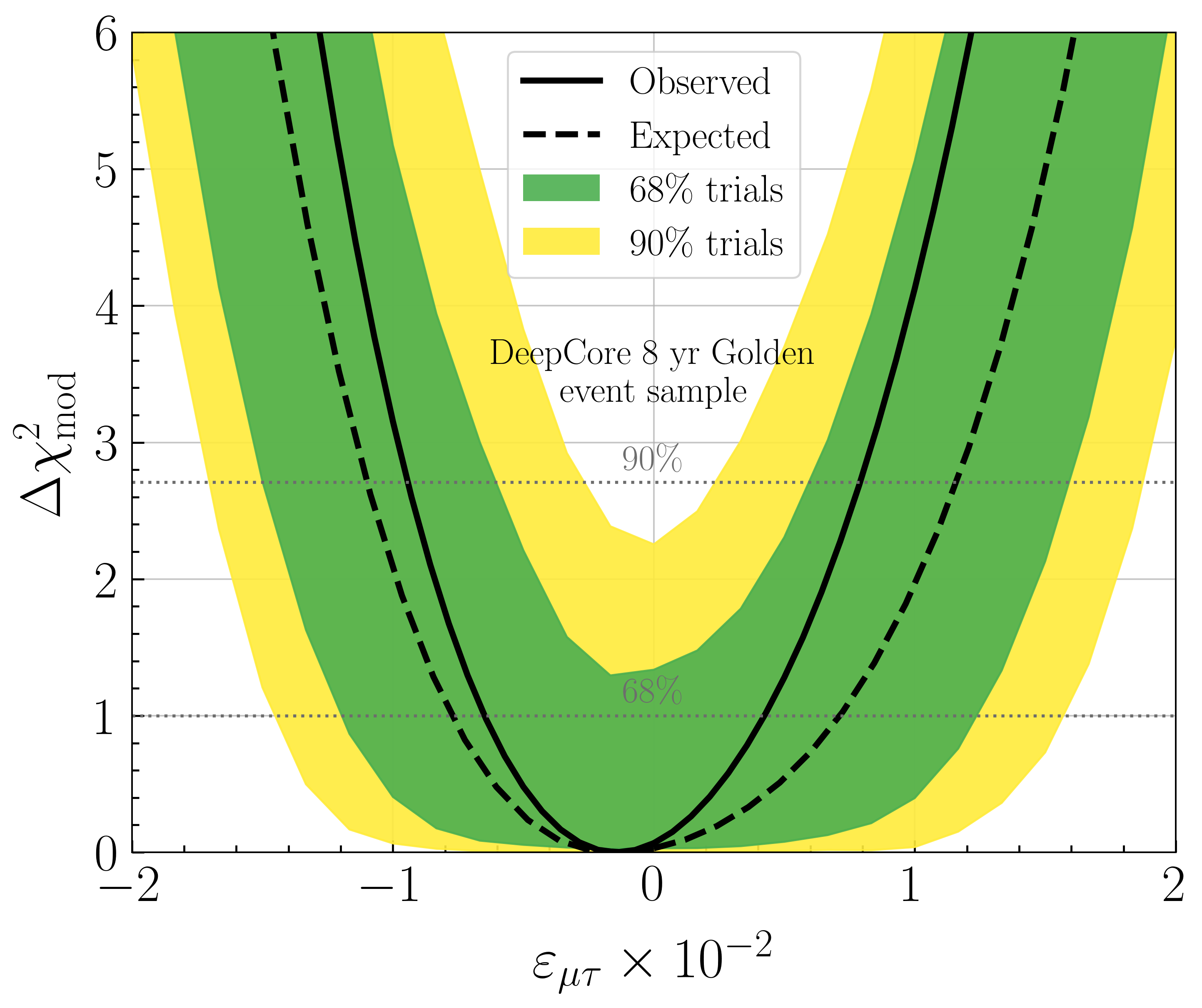}
		\caption{The observed (black solid curve) and the expected (black dashed curve) $\Delta\chi^2_{\rm mod}$ as a function of the $\varepsilon_{\mu\tau}$ parameter, assuming real values with the complex phase fixed to $0$ and $\pi$. We use the eight-year golden event sample of DeepCore. Shaded bands show the 68\% and 90\% ranges for the expected $\Delta\chi^2_{\rm mod}$ calculated from 500 statistically fluctuated pseudoexperiments. The horizontal dotted lines represent the 68\% and 90\% confidence levels for 1 DOF.}
		\label{fig:1d_scan}
	\end{figure}
	%%%%%%%%%%%%%%%%%%%%%%%%%%%%%%%%%%%
	
	To compare the constraints on the NSI parameter $\varepsilon_{\mu\tau}$ obtained from this analysis with those from other experiments (shown in Fig.~\ref{fig:nsi_comparison}), the parameter has been scanned assuming real values by fixing the phase $\phi_{\mu\tau}$ to 0 (\ie, positive $\varepsilon_{\mu\tau}$) and $\pi$ (\ie, negative $\varepsilon_{\mu\tau}$). Figure~\ref{fig:1d_scan} shows the observed (black solid curve) and the expected (black dashed curve) $\Delta\chi^2_{\rm mod}$ profiles for the real $\varepsilon_{\mu\tau}$. \textit{The 90\% CL allowed range is found to be $[-~0.0094, 0.0079]$ which is the most stringent bound obtained from any neutrino oscillation experiment to date.} The observed and the expected curves are consistent with $\Delta\chi^2$ distributions (bands) estimated from the statistically fluctuated pseudoexperiments within the $1\sigma$ band. Figure~\ref{fig:mutau_fc} in Appendix~\ref{app:fc_correction} shows the Feldman-Cousins~\cite{Feldman:1997qc} corrected 68\% confidence level, indicating that the limits reported here using the Wilks' theorem are valid and conservative.
	
	\begin{figure}[htp!]
		\centering
		\includegraphics[width=\linewidth]{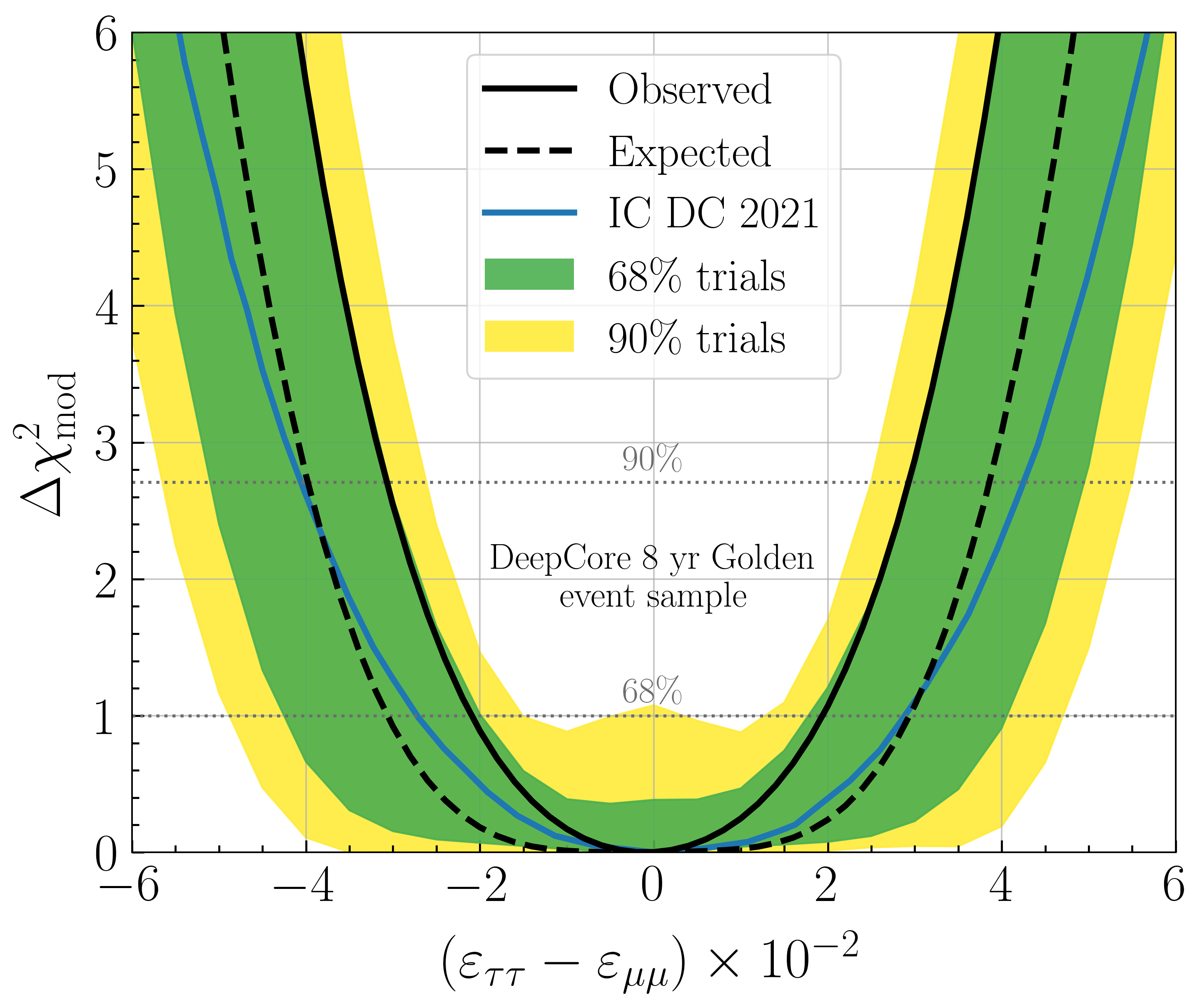}
		\caption{The observed (black solid curve) and the expected (black dashed curve) $\Delta \chi^2$ scan of NSI parameter $\varepsilon_{\tau\tau} - \varepsilon_{\mu\mu}$ using the eight-year golden event sample of DeepCore. Shaded bands show the 68\% and 90\% ranges of the expected $\Delta\chi^2$ calculated from 500 statistically fluctuated pseudoexperiments. The blue curve represents the $\Delta \chi^2$ from the previous IceCube DeepCore result~\cite{IceCubeCollaboration:2021euf}. The horizontal dotted lines represent the 68\% and 90\% confidence levels for 1 DOF.}
		\label{fig:tau_tau_result}
	\end{figure}
	
	Figure~\ref{fig:tau_tau_result} shows the observed (black solid curve) and the expected (black dashed curve) $\Delta\chi^2_{\rm mod}$ profiles for the nonuniversal NSI parameter $\varepsilon_{\tau\tau} - \varepsilon_{\mu\mu}$. The best-fit value of $\varepsilon_{\tau\tau} - \varepsilon_{\mu\mu}$ is found to be $-0.0009$, with a $\Delta\chi^2_{\rm mod}$ of 0.0017 relative to the standard interaction scenario. This indicates that the observed data are fully compatible with the standard interaction hypothesis and show no evidence for NSIs. The $p$-value of the NSI fit is 0.25, indicating good consistency of MC with the data (see Fig.~\ref{fig:data-mc} in the Appendix~\ref{app:data_mc}). \textit{The 90\% CL allowed range from the observed $\Delta\chi^2_{\rm mod}$ profiles (black solid curve) is $[-\,0.031, 0.029]$, which represents the strong bound under the assumption of NO.} The blue solid curve represents the $\Delta\chi^2_{\rm mod}$ profile reported in the previous IceCube DeepCore analysis~\cite{IceCubeCollaboration:2021euf}, demonstrating that the present result is consistent with the earlier measurement, while providing a stronger constraint. The observed $\Delta\chi^2_{\rm mod}$ profile is consistent with distributions of $\Delta\chi^2_{\rm mod}$ (bands) obtained from statistically fluctuated pseudoexperiments within the $1\sigma$ band. Figure~\ref{fig:tautau_fc} in Appendix~\ref{app:fc_correction} shows the Feldman-Cousins~\cite{Feldman:1997qc} corrected 68\% confidence level, indicating that the limits reported here using the Wilks' theorem are valid and conservative.\\
	%%%%%%%%%%%%%%%%%%%%%%%%%%%%%%%

	%%%%%%%%%%%%%%%%%%%%%%%%%%%%%%%%%%%%%%%%%%%%%%%%%%%%%%%%%%%%%%%%%%%%%%%
	\section{Conclusion}
	In this work, we use the publicly available high-purity $\nu_\mu$ CC  golden event sample collected during 2011--2019 by IceCube DeepCore to probe the NC-NSI parameters $\varepsilon_{\mu\tau}$ and $\varepsilon_{\tau\tau} - \varepsilon_{\mu\mu}$, which significantly affect the $\nu_\mu$ disappearance channel. Our analysis shows that the dataset is consistent with the standard interaction hypothesis with no evidence for NSI. Exploiting the excellent sensitivity of this data sample to $\nu_\mu$ disappearance channel, \textit{our analysis provides the  stringent bounds on $\varepsilon_{\mu\tau}$ and $\varepsilon_{\tau\tau} - \varepsilon_{\mu\mu}$} \textit{under the assumption of NO}. Despite operating in the challenging conditions of the Antarctic ice, the IceCube DeepCore detector has maintained its stable performance and continues to provide high-quality data, which helps us to place competitive constraints on these NSI parameters. The IceCube Upgrade~\cite{Ishihara:2019aao, IceCube:2025chb} currently being deployed is expected to further constrain these NSI parameters using its improved systematics, better reconstruction, and high statistics with a lower energy threshold.
	\\
	%%%%%%%%%%%%%%%%%%%%%%%%%%%%%%%%

	%%%%%%%%%%%%%%%%%%%%%%%%%%%%%%%%%%%%%%%%%%%%%%%%%%%%%%%%%%%%%%%%%%%%%%%
	\section{Acknowledgment}
	S.K.A., J.K., and A.K. acknowledge support from the Department of Atomic Energy (DAE), Govt. of India, and the Swarnajayanti Fellowship (Sanction Order No. DST/SJF/PSA-05/2019-20) provided by the Department of Science and Technology (DST), Govt. of India, and the Research Grant (Sanction Order No. SB/SJF/2020-21/21) provided by the Anusandhan National Research Foundation (ANRF), Govt. of India, under the Swarnajayanti Fellowship. We would like to express our sincere thanks and gratitude to Professor Francis Halzen, the 2026 Nobel laureate in Physics, for his constant support and encouragement throughout this study. The numerical simulations were performed using the Dell PowerEdge R660 Server and the “SAMKHYA: High-Performance Computing Facility” at the Institute of Physics,  Bhubaneswar, India.\\
	
	%%%%%%%%%%%%%%%%%%%%%%%%%%%%%%%%
	\section{Data availability}
	In this work, we analyze the publicly available data from IceCube Collaboration~\cite{DVN_B4RITM_2025}. Constraints on the NSI parameters obtained in this study are available from the authors in the form of digitized files upon reasonable request.
	%%%%%%%%%%%%%%%%%%%%%%%%%%%%%%%%%%%%%%%%%%%%%%%%%%%%%%%%%%%%%%%%%%%%%%%%%%%%%%%

	%%%%%%%%%%%%%%%%%%%%%%%%%%%%%%%%%%%%%%%%%%%%%%%%%%%%%%%%%%%%%%%%%%%%%%%%%%%%%%%
	%%%%%%%%%%%%%%%%%%%%%%%%%%%%%%%%%%%%%%%%%%%%%%%%%%%%%%%%%%%%%%%%%%%%%%%%%%%%%%%
	
%	\newpage
%	\clearpage
	
	\appendix
	
	%%%%%%%%%%%%%%%%%%%%%%%%%%%%%%%%%%%%%%%%%%%%%%%%%%%%%%%%%%%%%%%%%%%%%%%%%%%%%%%
	%%%%%%%%%%%%%%%%%%%%%%%%%%%%%%%%%%%%%%%%%%%%%%%%%%%%%%%%%%%%%%%%%%%%%%%%%%%%%%%
	
%	\onecolumngrid
	
%	\begin{center}
	%	\textbf{End Matter}
	%\end{center}
	
%	\twocolumngrid	
	
	%%%%%%%%%%%%%%%%%%%%%%%%%%%%%%%%%%%%%%%%%%%%%%%%%%%%%%%%%%%%%%%%%%%%%%%%%%%%%%%
	%%%%%%%%%%%%%%%%%%%%%%%%%%%%%%%%%%%%%%%%%%%%%%%%%%%%%%%%%%%%%%%%%%%%%%%%%%%%%%%
	
	\section{Systematic uncertainties}
	\label{app:systematic}
	
	This analysis follows the systematic uncertainty treatment described in Ref.~\cite{IceCubeCollaboration:2023wtb}, where we incorporate uncertainties related to the detector response, atmospheric neutrino flux, neutrino interaction cross section, normalization, and neutrino oscillations. Among the detector-related parameters, the uncertainty in the overall optical efficiency of the DOMs is represented by a single scaling parameter. The ice absorption and ice scattering parameters account for variations in the absorption and scattering lengths of the bulk ice. Two additional parameters, $p_0$ and $p_1$~\cite{IceCube:2023ahv}, describe the relative angular acceptance of the optical modules, accounting for changes in the optical properties of refrozen ice along each string column.
	
	Atmospheric neutrino flux predictions are based on the model from Ref.~\cite{Honda:2015fha}. The associated uncertainties include the spectral index parameter ($\Delta\gamma_{\nu}$) and uncertainties related to the production of pions and kaons, following the method provided in Ref.~\cite{Barr:2006it}. Uncertainties in the neutrino-nucleon interaction cross sections are parametrized through the axial mass parameters corresponding to the quasielastic (QE) and the resonant (RES) interactions. For deep inelastic scattering (DIS) interactions, an interpolated model between GENIE~\cite{GENIE-cross} and CSMS~\cite{Cooper-Sarkar:2011jtt} predictions is used. An additional normalization parameter is included to account for uncertainties in the NC-type event rate. Apart from these parameters, overall normalization parameters for the neutrinos and the atmospheric muon background are also included. In the set of neutrino oscillation parameters, the uncertainties from $\theta_{23}$ and $\Delta m^2_{31}$ are also considered in this work.
	
	The list of systematic uncertainty parameters, along with their corresponding nominal values, $1\sigma$ priors, and ranges, is summarized in Table~\ref{tab:systematic_params}. A pull penalty term is added to $\chi^2_{\rm mod}$ for the parameters where priors are available. Parameters without priors are assumed to follow uniform distributions with no pull penalty. The table also reports the best-fit values of the systematic parameters obtained from the two independent fits for SI + NSI scenarios--one for $\varepsilon_{\mu\tau}$ and another for $\varepsilon_{\tau\tau}-\varepsilon_{\mu\mu}$--taken one at a time, with the other NSI parameters considered to be zero. The best-fit values of these nuisance parameters are found to be consistent between the two fits.
	
	\onecolumngrid

	\begin{table}[ht]
		\centering
		\renewcommand{\arraystretch}{1.3}
		\begin{tabular}{l@{\hskip 15pt}c@{\hskip 15pt}c@{\hskip 15pt}c@{\hskip 15pt}c@{\hskip 15pt}c}
			\hline
			\hline
			\textbf{Parameter} & \textbf{Best fit} ($\varepsilon_{\mu\tau}$) & \textbf{Best fit} ($\varepsilon_{\tau\tau}-\varepsilon_{\mu\mu}$) &\textbf{Nominal value} & \textbf{Prior ($1\sigma$)} & \textbf{Range} \\
			\hline
			\multicolumn{6}{@{}l}{\textbf{Detector:}} \\
			DOM efficiency		     & 1.064 	& 1.064 		& 1.0 		& $\pm~$0.1 & [0.8, 1.2] \\
			Ice absorption           & 0.973 	& 0.973 		& 1.0 		& Unconstrained &[0.9, 1.1] \\
			Ice scattering           & 0.987 	& 0.987 		& 1.05 		& Unconstrained &[0.95, 1.15] \\
			Relative eff. $p_0$      & $-\,0.271$ 	& $-\,0.269$ 	& 0.10 		& Unconstrained &[$-\,0.2$, 0.6] \\
			Relative eff. $p_1$      & $-\,0.043$ & $-\,0.042$ 	& $-\,0.05$ 	& Unconstrained &[$-\,0.2$, 0.2] \\
			\hline
			\multicolumn{6}{@{}l}{\textbf{Atmospheric neutrino flux:}} \\
			$\Delta \gamma_{\nu}$     & 0.063 & 0.063 & 0.0 & $\pm$\,0.1 & [$-\,0.5$, 0.5] \\
			$\Delta \pi^\pm \text{ yields [A-F]}$ & 0.06 & 0.06 & 0.0 & $\pm\,$0.3 & [$-\,1.5$, 1.5] \\
			$\Delta \pi^\pm \text{ yields G}$& $-\,0.048$ & $-\,0.05$ & 0.0 & $\pm\,$0.3 & [$-\,1.5$, 1.5] \\
			$\Delta \pi^\pm \text{ yields H}$& $-\,0.013$ & $-\,0.017$ & 0.0 & $\pm\,$0.15 & [$-\,0.75$, 0.75] \\
			$\Delta K^+ \text{ yields W}$ & 0.078 & 0.084 & 0.0 & $\pm\,$0.4 & [$-\,2.0$, 2.0]\\
			$\Delta K^+ \text{ yields Y}$ & 0.095 & 0.107 & 0.0 & $\pm\,$0.3 & [$-\,1.5$, 1.5]\\
			$\Delta K^- \text{ yields W}$ & $-\,0.012$ & $-\,0.009$ & 0.0 & $\pm\,$0.4 & [$-\,2.0$, 2.0]\\
			\hline
			\multicolumn{6}{@{}l}{\textbf{Neutrino interaction cross section:}} \\
			$M_A^{\text{CCQE}}$ (in $\sigma$)  & 0.058 & 0.061  & 0.0 & $\pm\,$1.0 & [$-\,2.0$, 2.0]\\
			$M_A^{\text{CCRES}}$ (in $\sigma$)& 0.60 & 0.606 & 0.0 & $\pm\,$1.0 & [$-\,2.0$, 2.0]\\
			DIS CSMS                & 0.044 & 0.033 & 0.0 & $\pm\,$1.0 & [$-\,3.0$, 3.0]\\
			$\sigma_{\rm NC}/\sigma_{\rm CC} $ & 1.129 & 1.127 & 1.0 & $\pm\,$0.2 & [0.5, 1.5]\\
			\hline
			\multicolumn{6}{@{}l}{\textbf{Normalization:}} \\
			$A_{\text{eff}}$ scale   & 0.823 & 0.824 & 1.0 & Unconstrained &[0.6, 1.4] \\
			\hline
			\multicolumn{6}{@{}l}{\textbf{Atmospheric muons:}} \\
			Atm. $\mu$ scale         & 1.349 & 1.364 & 1.0 & Unconstrained &[0.7, 1.5] \\
			\hline
			\multicolumn{6}{@{}l}{\textbf{Oscillations:}} \\
			$\theta_{23}$           & 45.343$^\circ$ & 45.326$^\circ$ & 45.573$^\circ$ & Unconstrained &[38$^\circ$, 52$^\circ$] \\
			$\Delta m^2_{31}$       & 0.002498 eV$^2$ & 0.002488 eV$^2$ & 0.002484 eV$^2$ & Unconstrained &[0.002, 0.003] eV$^2$ \\
			\hline
			\hline
		\end{tabular}
		
		\caption{The table shows the systematic uncertainty parameters considered as nuisance parameters in this analysis. The nominal value, associated $1\sigma$ prior (if available), and the allowed range during minimization are mentioned for each parameter. The table also includes the best-fit values for these parameters obtained from the data fitting with the SI + NSI hypothesis having parameters $\varepsilon_{\mu\tau}$ (second column) and $\varepsilon_{\tau\tau}-\varepsilon_{\mu\mu}$ (third column) considered one at a time.}
		\label{tab:systematic_params}
	\end{table}
	\newpage
	\twocolumngrid

	\section{Effects of NSI on DeepCore events}
	\label{app:event_diff}

	\begin{figure}[htp!]
		\includegraphics[width=\linewidth]{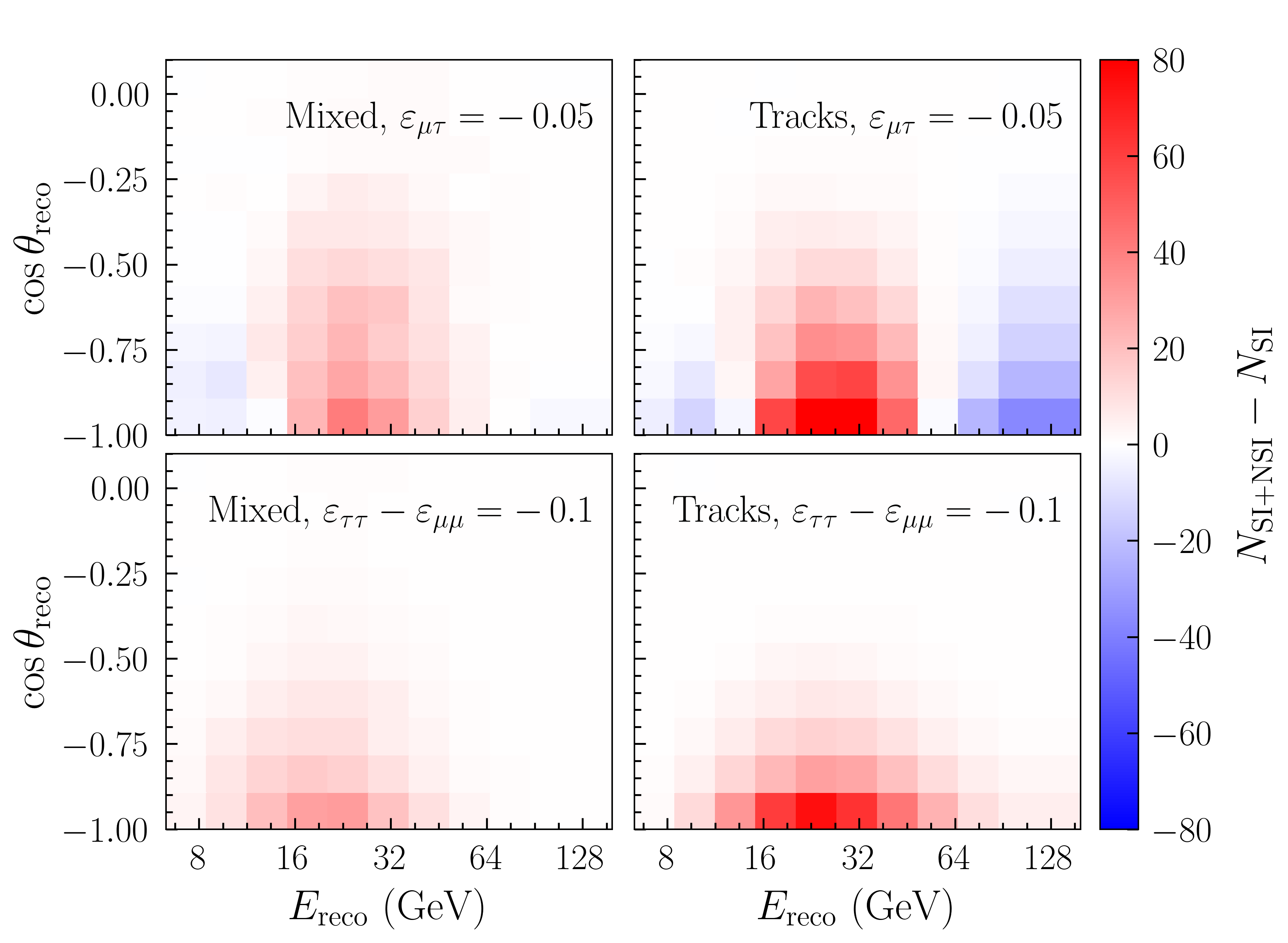}
		\caption{Difference of the expected event distributions between SI + NSI and SI scenarios (\ie, $N_{\text{SI+NSI}} - N_{\text{SI}}$) in the plane of the reconstructed energy $E_{\text{reco}}$ and the reconstructed cosine of zenith angle $\cos\theta_{\text{reco}}$, for two NSI parameter choices taken one at a time. The top and bottom panels correspond to $\varepsilon_{\mu\tau} = -\,0.05$ and $\varepsilon_{\tau\tau} - \varepsilon_{\mu\mu} = -\,0.1$, respectively. The left and right panels show distributions for mixed and tracklike samples, respectively. Here, we use the nominal values of the nuisance parameters, as given in Table~\ref{tab:systematic_params}.
		}
		\label{fig:event_dif_neg}
	\end{figure}
	
	Figure~\ref{fig:event_dif_neg} shows the distributions for event differences between the SI + NSI and the SI scenarios with the benchmark values of $\varepsilon_{\mu\tau}=-\,0.05$ and $\varepsilon_{\tau\tau}-\varepsilon_{\mu\mu}=-\,0.1$. The distributions of event differences highlight the regions in reconstructed energy and cosine zenith that contribute significantly to the sensitivity for NSI. The top panels corresponding to $\varepsilon_{\mu\tau}=-\,0.05$ show the event differences with alternating signs in the region of the oscillation valley up to the high-energy region, while the bottom panels corresponding to $\varepsilon_{\tau\tau}-\varepsilon_{\mu\mu}=-\,0.1$ show event differences only in the positive direction below 100 GeV. From the comparison between Fig.~\ref{fig:event_dif_pos} and Fig.~\ref{fig:event_dif_neg}, it can be observed that the shape of the signal region in the distribution of event difference changes with the sign of $\varepsilon_{\mu\tau}$. In contrast, for $\varepsilon_{\tau\tau}-\varepsilon_{\mu\mu}$, the signal region remains nearly identical irrespective of the sign of the parameter.
	
	\section{Data-MC Comparison}
	\label{app:data_mc}
	
\begin{figure}[htp!]
		\includegraphics[width=\linewidth]{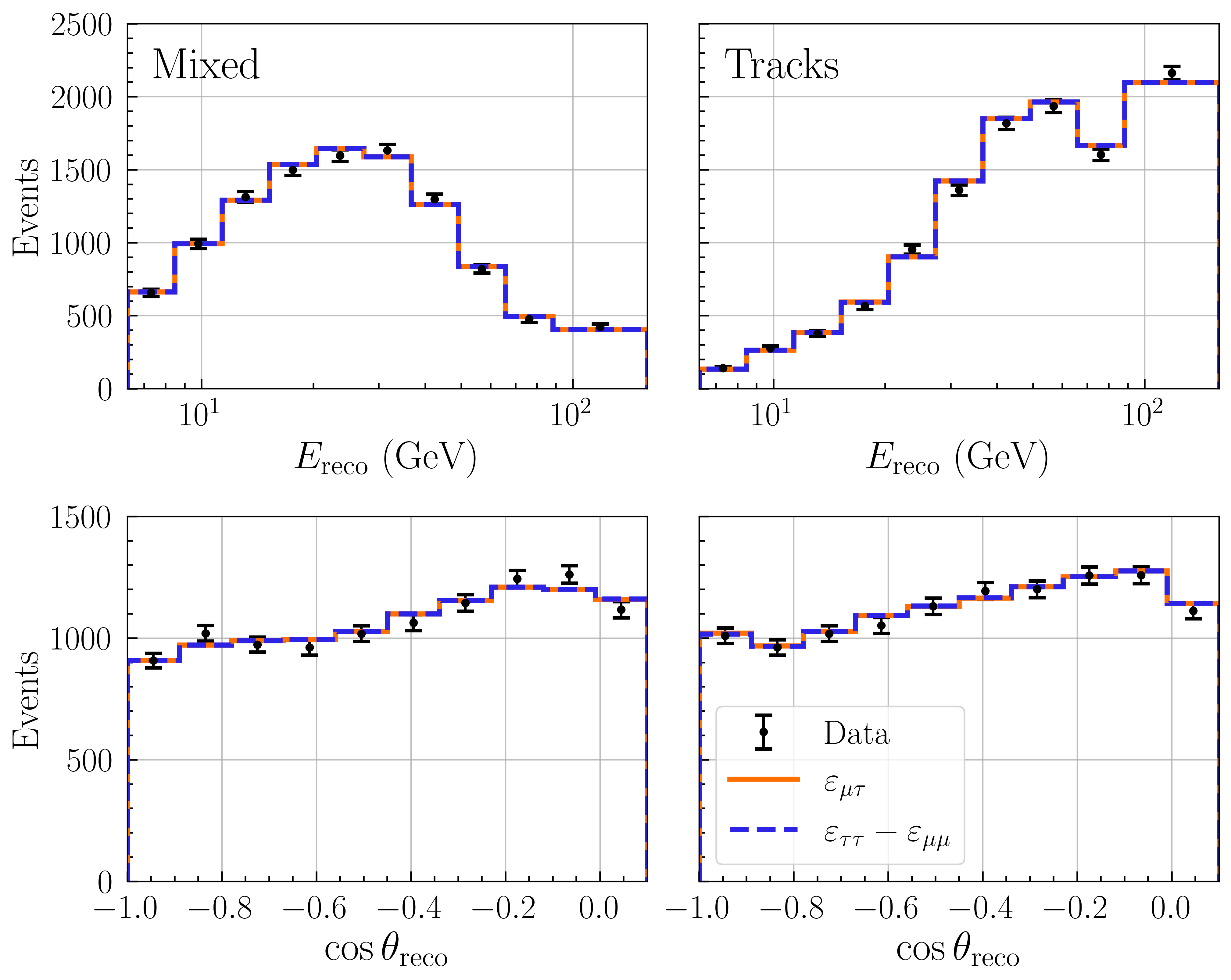}
		\caption{Comparison of reconstructed event distributions between the observed data in the eight-year golden event sample of DeepCore and the best-fit MC predictions for different NSI scenarios as a function of energy $E_{\text{reco}}$ (top panels) and the cosine of zenith angle $\cos\theta_{\text{reco}}$ (bottom panels). The left and right panels correspond to the mixed and tracklike event samples, respectively. The black dots denote the data with statistical uncertainties. The orange solid and blue dashed lines represent the best-fit MC predictions for the NSI scenarios with $\varepsilon_{\mu\tau}$ and $\varepsilon_{\tau\tau} - \varepsilon_{\mu\mu}$, respectively, taken one at a time.}
		\label{fig:data-mc}
	\end{figure}

	Figure~\ref{fig:data-mc} shows the comparison of the 1D projections of the observed data and the best-fit MC predictions for the NSI parameters $\varepsilon_{\mu\tau}$ and $\varepsilon_{\tau\tau}-\varepsilon_{\mu\mu}$ taken one at a time. The event distributions are shown as functions of the reconstructed energy (top panels) and the cosine zenith (bottom panels), separately for each PID. The best-fit MC predictions obtained from the independent fits for SI + NSI scenarios with $\varepsilon_{\mu\tau}$ and $\varepsilon_{\tau\tau}-\varepsilon_{\mu\mu}$ taken one at a time show good agreement with the observed data.
	
	\section{Impact of neutrino mass ordering uncertainties on NSI sensitivities}
	\label{app:io_sensitivity}
	
\begin{figure}[htp!]
		\includegraphics[width=\linewidth]{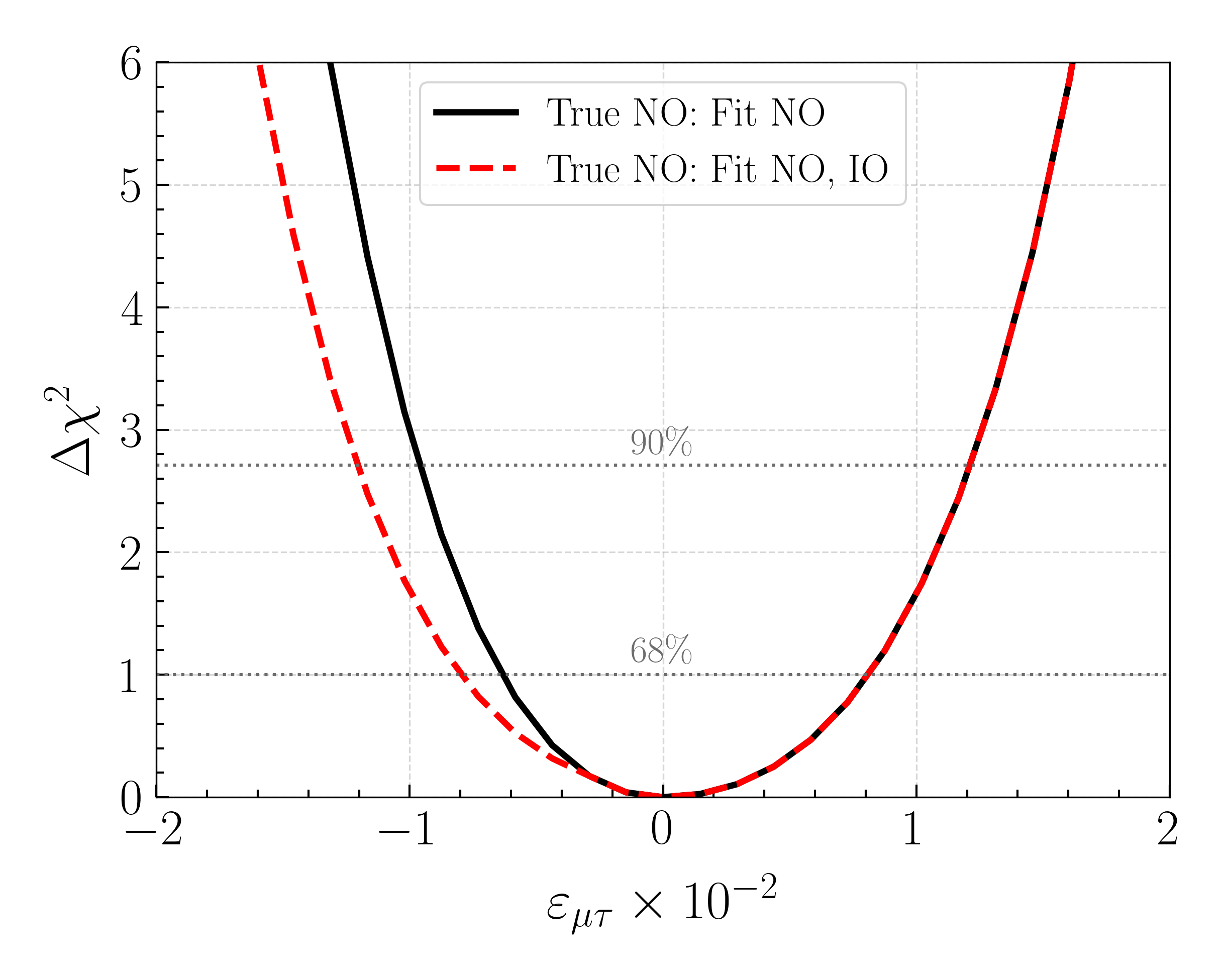}
		\caption{Sensitivity of $\varepsilon_{\mu\tau}$ assuming real values, with the complex phase fixed to $0$ and $\pi$, using the MC equivalent of the eight-year DeepCore golden event sample with NO as the true mass ordering. The sensitivities obtained with (red dashed curve) and without (black solid curve) minimizing over the neutrino mass ordering are compared. Detector-related systematics are excluded from the sensitivity calculation (see the text for details).}
		\label{fig:mutau_io_sensitivity}
	\end{figure}

\begin{figure}[htp!]
		\includegraphics[width=\linewidth]{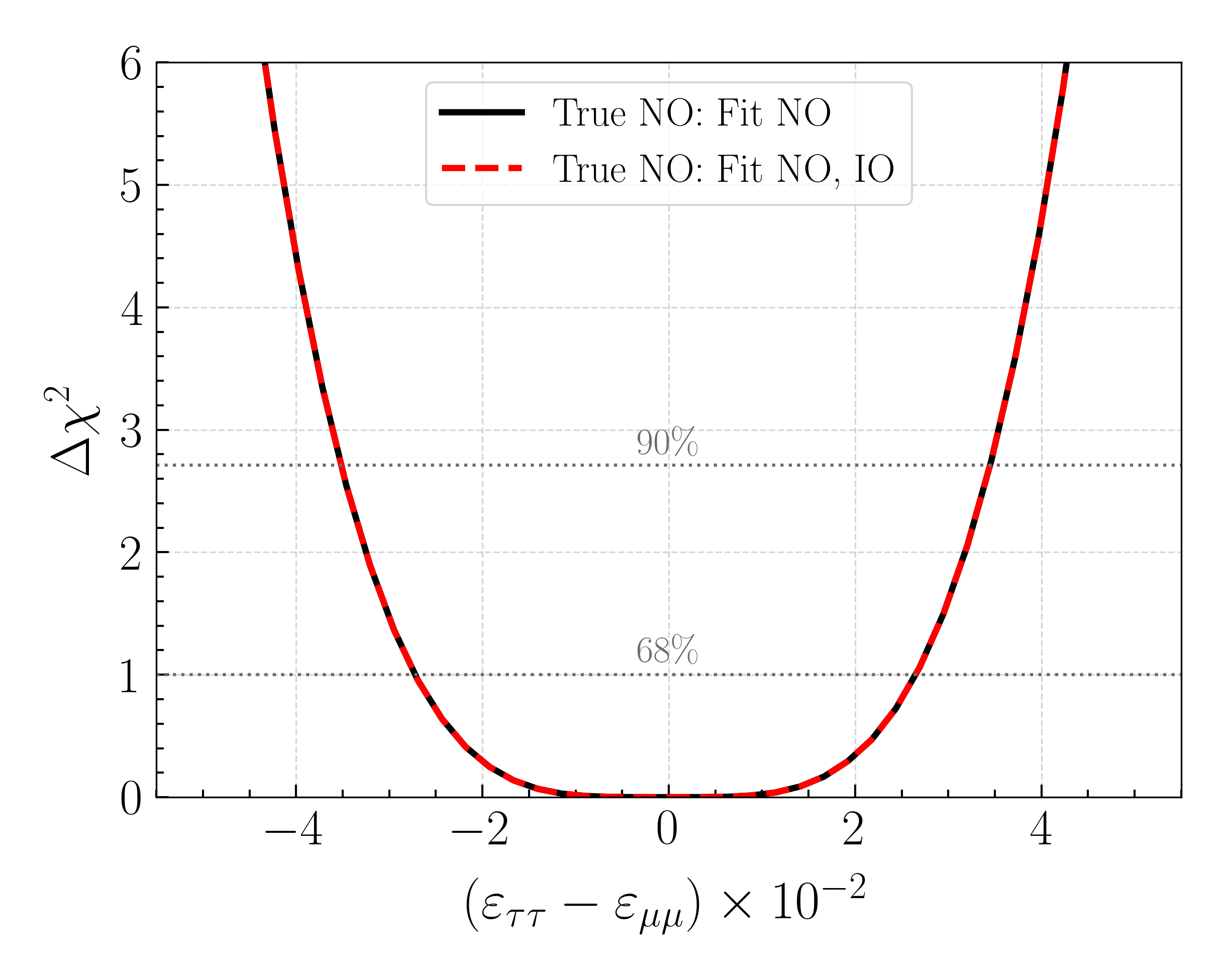}
		\caption{Sensitivity of $\varepsilon_{\tau\tau}-\varepsilon_{\mu\mu}$ using the MC equivalent of the eight-year DeepCore golden event sample with NO as the true mass ordering. The sensitivities obtained with (red dashed curve) and without (black solid curve) minimizing over the neutrino mass ordering are compared. The detector-related systematics are excluded in this sensitivity calculation (see the text for a detailed explanation).}
		\label{fig:tautau_io_sensitivity}
		
	\end{figure}

	In this appendix, to assess the impact of the neutrino mass ordering uncertainty, we compare the NSI sensitivities obtained with and without minimizing over the neutrino mass ordering. The detector-related systematic treatment in this analysis is handled by the ``hypersurface'' method, which is described in detail in Ref.~\cite{IceCubeCollaboration:2023wtb}. The hypersurface depends on the value of the oscillation parameter $\Delta m^{2}_{31}$ because of its correlation with the detector systematics. Therefore, hypersurfaces have to be generated separately for NO and IO. However, in the public data release, the hypersurface is only provided for NO. Due to this limitation, it is not possible to incorporate the detector-related systematics when minimizing over the mass ordering. Therefore, we exclude the detector-related systematics in this comparison. Figures~\ref{fig:mutau_io_sensitivity} and~\ref{fig:tautau_io_sensitivity} show the comparison of the sensitivities for the NSI parameters $\varepsilon_{\mu\tau}$ and $\varepsilon_{\tau\tau}-\varepsilon_{\mu\mu}$, respectively, assuming NO as the true mass ordering, without considering the detector-related systematics. In these figures, we observe comparable NSI sensitivities with and without minimizing over the neutrino mass ordering. This indicates that the impact of minimization over the mass ordering on the NSI sensitivities is small. This is expected because, irrespective of the neutrino mass ordering, both neutrinos and antineutrinos are affected by NSI, unlike the standard matter effect, which is felt only by neutrinos when the ordering is normal and by antineutrinos when the ordering is inverted.
	
	%\newpage
	%\clearpage

	\section{Validity of the Wilks' theorem: Feldman-Cousins correction}
	\label{app:fc_correction}
	
\begin{figure}
		\includegraphics[width=\linewidth]{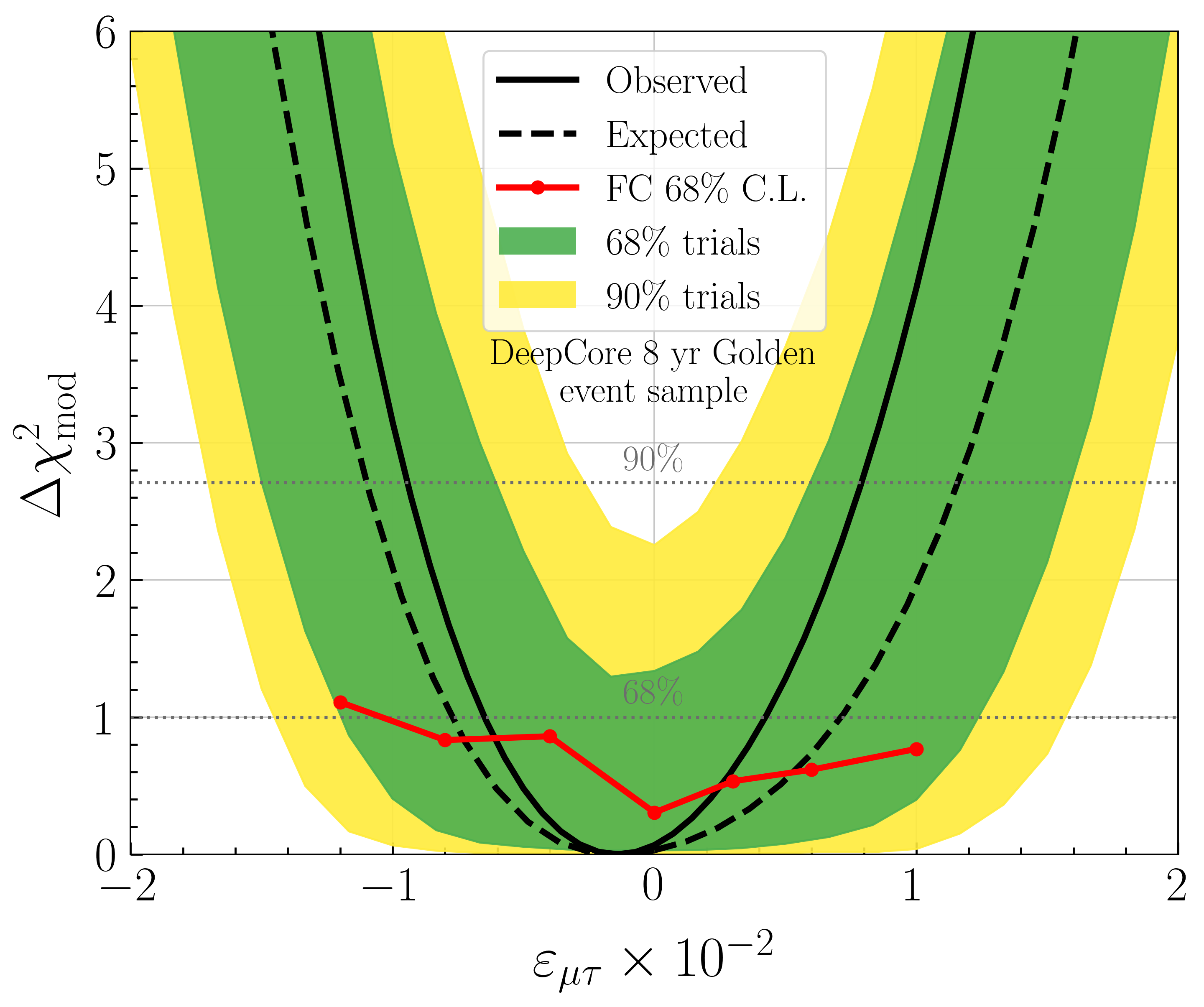}
		\caption{The observed (black solid curve) and the expected (black dashed curve) $\Delta\chi^2_{\rm mod}$ as a function of the $\varepsilon_{\mu\tau}$ parameter, assuming real values with the complex phase fixed to $0$ and $\pi$. We use the eight-year golden event sample of DeepCore. Shaded bands show the 68\% and 90\% ranges for the expected $\Delta\chi^2_{\rm mod}$ calculated from 500 statistically fluctuated pseudoexperiments. The horizontal dotted lines represent the 68\% and 90\% confidence levels for 1 DOF, following Wilks' theorem. The red solid curve represents the FC-corrected 68\% confidence level.}
		\label{fig:mutau_fc}
	\end{figure}

\begin{figure}
		\includegraphics[width=\linewidth]{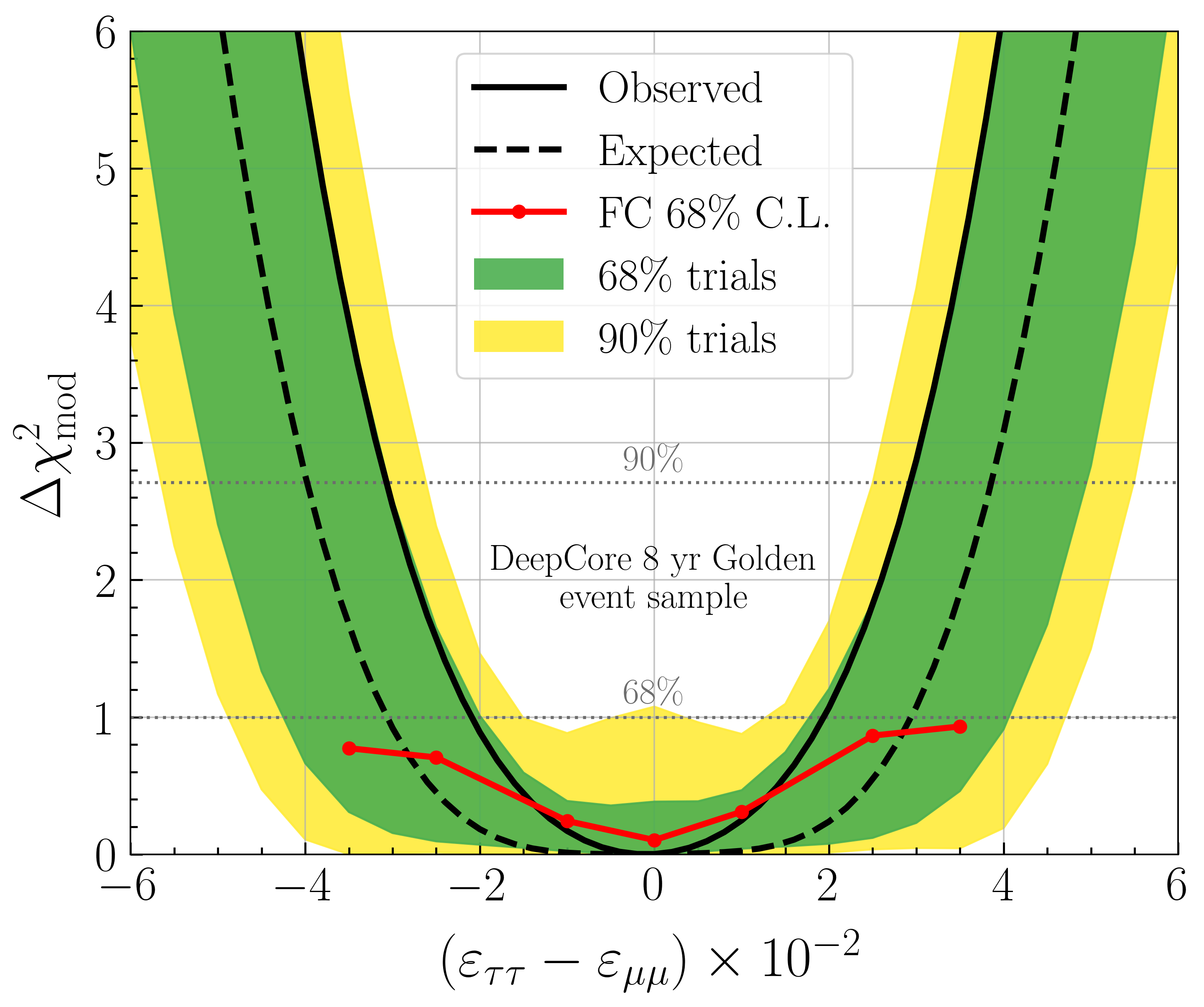}
		\caption{The observed (black solid curve) and the expected (black dashed curve) $\Delta \chi^2$ scan of NSI parameter $\varepsilon_{\tau\tau} - \varepsilon_{\mu\mu}$ using the eight-year golden event sample of DeepCore. Shaded bands show the 68\% and 90\% ranges of the expected $\Delta\chi^2$ calculated from 500 statistically fluctuated pseudoexperiments. The horizontal dotted lines represent the 68\% and 90\% confidence levels for 1 DOF, following Wilks' theorem. The red solid curve represents the FC-corrected 68\% confidence level.}
		\label{fig:tautau_fc}
	\end{figure}

	The Feldman-Cousins (FC) correction~\cite{Feldman:1997qc} is performed at a few representative values of the NSI parameters, $\varepsilon_{\mu\tau}$ and $\varepsilon_{\tau\tau}-\varepsilon_{\mu\mu}$,  to verify the validity of the Wilks' theorem. Figures~\ref{fig:mutau_fc} and~\ref{fig:tautau_fc} show the FC-corrected 68\% confidence levels (red curves) for the parameters $\varepsilon_{\mu\tau}$ and $\varepsilon_{\tau\tau}-\varepsilon_{\mu\mu}$, respectively. For this FC correction, we use the nominal values of the nuisance parameters (see Table~\ref{tab:systematic_params}) to generate the pseudoexperiments. Both Figs.~\ref{fig:mutau_fc} and~\ref{fig:tautau_fc} show that the largest deviation from the Wilks' expectation is observed around the NSI value of zero. As the true NSI values move away from zero, the FC-corrected thresholds approach the Wilks' expectation. Overall, the FC study demonstrates that the limits obtained using Wilks' theorem are valid and conservative.

\newpage
%%%%%%%%%%%%%%%%%%%%%%%%
%\bibliography{references.bib}


\end{document}